\documentclass[journal]{IEEEtran}
\usepackage{amsmath, amsthm, amssymb, amsfonts}
\usepackage{mathtools}
\usepackage{bm}
\usepackage{graphicx}
\usepackage{url}

\usepackage{atbegshi}
\AtBeginDocument{\AtBeginShipoutNext{\AtBeginShipoutDiscard}}

\begin{document}
\title{The design and performance of the real time software architecture for the ITER Radial Neutron Camera}
\author{N. Cruz,~\IEEEmembership{Member,~IEEE}, B. Santos, A. Fernandes, P.F. Carvalho, J. Sousa, B. Gon\c{c}alves, M. Riva, C. Centioli, D. Marocco, B. Esposito, C.M.B.Correia, and R.C. Pereira}
\thanks{N. Cruz (email: nunocruz@ipfn.tecnico.ulisboa.pt), B. Santos,	A. Fernandes,	P.F. Carvalho, J. Sousa, B. Gon\c{c}alves, and R.C. Pereira are with Instituto de Plasmas e Fus\~ao Nuclear Instituto Superior T\'ecnico Universidade de Lisboa, 1049-001 Lisboa, Portugal.}
\thanks{M. Riva, C. Centioli, D. Marocco and B. Esposito are with ENEA C. R. Frascati, Dipartimento FSN, via E. Fermi 45, 00044 Frascati (Roma), Italy.}
\thanks{C.M.B.Correia is with LIBPhys-UC, Department of Physics, University of Coimbra, P-3004 516 Coimbra, Portugal.}
\markboth{The design and performance of the real time
software architecture for the ITER Radial Neutron
Camera}%
{Shell \MakeLowercase{\textit{et al.}}: Bare Demo of IEEEtran.cls for IEEE Journals}
\maketitle

\begin{abstract}
The neutron detection system for characterization of emissivity in ITER Tokamak during DD and DT experiments poses serious challenges to the performance of the diagnostic control and data acquisition system (CDAcq). The ongoing design of the ITER Radial Neutron Camera (RNC) diagnostic is composed by 26 lines of sight (LOS) for complete plasma inspection. The CDAcq system aims at meeting the ITER requirements of delivering the measurement of the real-time neutron emissivity profile with time resolution and control cycle time of 10 ms at peak event rate of 2 MEvents/s per LOS. This measurement demands the generation of the neutron spectra for each LOS with neutron/gamma discrimination and pile up rejection. The neutron spectra can be totally processed in the host CPU or it can use the processed data coming from the system FPGA \cite{IEEERT2018:Ana}. The number of neutron counts extracted from the spectra is then used to calculate the neutron emissivity profile using an inversion algorithm. Moreover, it is required that the event based raw data acquired is made available to the ITER data network without local data storage for post processing. The data production for the 2 MEvents/s rate can go up to a maximum data throughput of 0.5 GB/s per channel, fostering the evaluation of  real time data compression techniques in RNC \cite{IEEERT2018:Bruno}. 

To meet the demands of the project a CDAcq prototype has been used to design and test a high-performance distributed software architecture taking advantage of multi-core CPU technology capable of coping with the requirements. This submission depicts the design of the real-time architecture, the spectra algorithms (pulse height analysis, neutron/gamma discrimination and pile-up correction) and the inversion algorithm to calculate the emissivity profile. Preliminary tests to evaluate the system’s performance with synthetic data are presented. 
\end{abstract}


%
\IEEEpeerreviewmaketitle

\section{Introduction}
\IEEEPARstart{T}{he} Radial Neutron Camera (RNC) is an ITER diagnostic that aims to deliver in real time with a 10 ms control cycle the profile of the plasma neutron emission for machine control purposes. Each line of sight (LOS) provides the integrated measurement of the neutron flux by means of spectrometers and neutron flux monitors. These measurements provide the inputs for the neutron emissivity calculation using an inversion algorithm that uses the magnetic surfaces and the integrated neutron flux \cite{Marocco_2011}\cite{RNC_Marocco_2012}\cite{RNC_Cruz_2017}.  

\begin{figure}
	\centering
	\includegraphics[width=8.8cm]{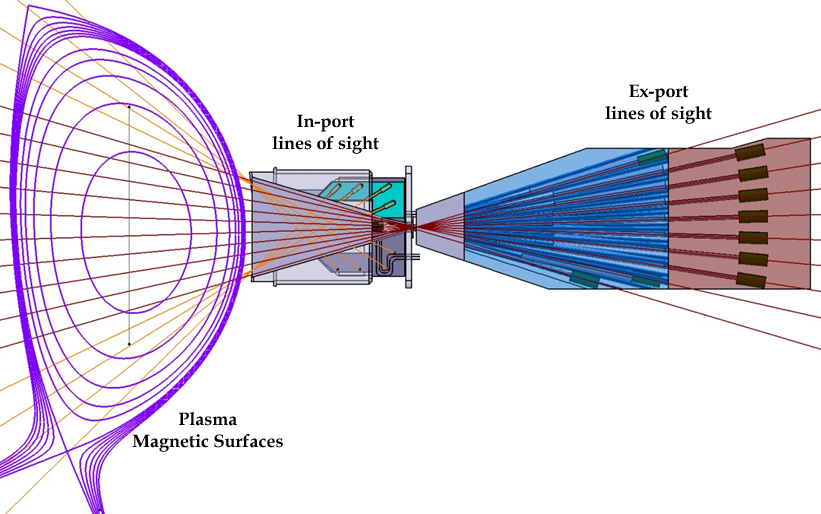}
	\caption{RNC diagram with the lines of sight and the plasma magnetic surfaces.}
	\label{fig:RNC_LOS}
\end{figure}

Figure \ref{fig:RNC_LOS} depicts the system architecture with the external and internal port LOS and the plasma magnetic surfaces used to calculate the neutron emissivity in the points of intersection between the LOS and the magnetic surfaces.  

The design of the high-performance distributed software architecture using multi-core CPU technology capable of coping with the system highly demanding requirements was validated using a control and data acquisition prototype. This contribution describes the real time architecture, with the spectrum neutron and gamma energy spectrum construction algorithms (including pulse height analysis, neutron/gamma discrimination and pile-up correction) and the inversion algorithm to calculate the neutron emissivity profile. The description of the preliminary tests made to validate and evaluate the system’s performance with synthetic data are presented.

\section{Overall System Architecture}

To obtain the necessary measurements that comply with the system requirements, the ITER RNC diagnostic runs a set of functions implemented in a variety of algorithms performing the following tasks: 
\begin{itemize}
	\item acquisition of streamed  data from radiation detectors; 
	\item detection of pulses (events) to build a sequence of relevant data for the measurement, while discarding the stream where no pulses were detected;
	\item  pulse processing in order to provide count rates and pulse height spectra; 
	\item calculation of the neutron emissivity profile reconstruction (by processing the neutron count rate data together with the magnetic flux information). 
\end{itemize}

While the neutron emissivity profile real time algorithms must run within 10 ms cycles providing the measurement in every cycle to be used for advanced control purposes, the stream of  raw detector pulses data must be delivered for storage aiming at offline physics studies. For this purpose data compression algorithms were implemented aiming at reducing the size of detector pulses data before they are sent to the ITER archiving system. 

\section{System Prototype}
This section describes the control and data acquisition prototype designed to test the multi-core CPU technology distributed software architecture, the implemented algorithms and the performance of the solutions presented to meet the demanding RNC requirements. 

Although ATCA \cite{ATCA_ITER_FCS} or PXIe \cite{PXIe_ITER_FCS} architectures may be used for the final system design, the decision is still pending, based on the performance test results, cubicle space optimization and reduction of unnecessary costs. Thus, the prototype architecture (PCIe bus) was decided based on a cost effective solution capable of deploying the necessary performance tests, software architecture tools and critical algorithms \cite{RNC_Pereira_2017}.

The test and development prototype hardware and software specifications feature:
\begin{itemize}
	\item Real-time OS: Scientific Linux 7.0 with RT kernel;	
	\item In-house developed device drivers;	
	\item Host PC running on Intel(R) Core(TM) i7-5930K CPU @ 3.50GHz with 6 cores and 12 independent software threads, 64 GB memory and 256 GB SSD;
	\item Xilinx KC705 evaluation board featuring a Kintex 7 FPGA;
  \item In-house developed firmware featuring ADC readout, PCIe DMA data streaming management, event detection and pulse analysis \cite{IEEERT2018:Ana}; 
	\item In-house developed analogue input with 2 ADC 12-bit @ up to 1.6 GHz FPGA Mezzanine Card (FMC-AD2-1600) \cite{RNC_Pereira_2017}.
\end{itemize}
	
\begin{figure}
	\centering
	\includegraphics[width=5.55cm]{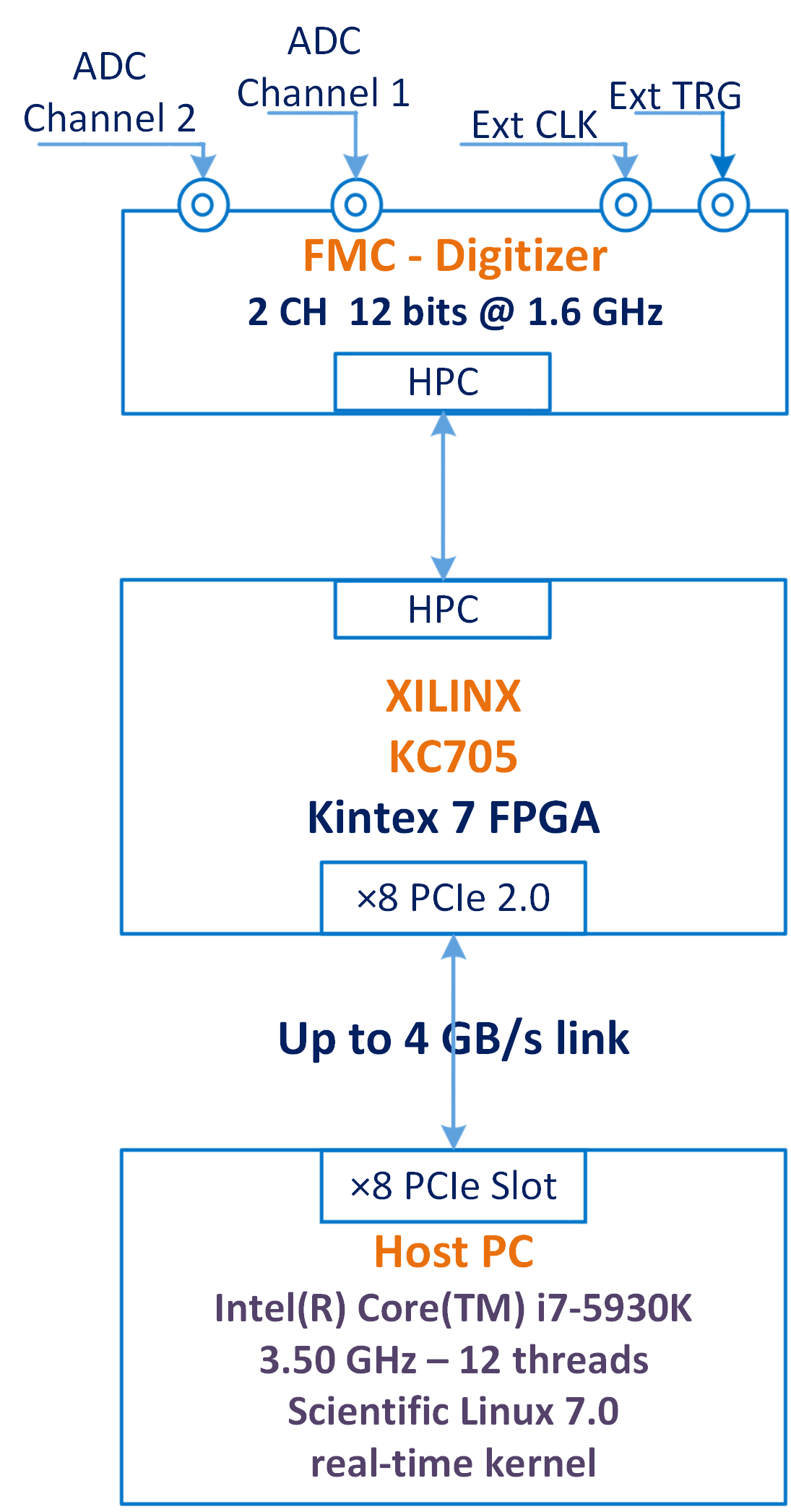}
	\caption{RNC Data Acquisition and Processing Prototype Architecture}
	\label{fig:Prototype_Architecture}
\end{figure}

Figure \ref{fig:Prototype_Architecture} depicts the prototype hardware setup with one FMC digitizer. The system can also host another Xilinx KC705 simultaneously increasing the number of analogue channels per host. Nevertheless the inclusion of more channels demand that the host has enough performance and CPU cores available to run all the algorithms. 

\section{Software Architecture}

\begin{figure*}
	\centering
	\includegraphics[width=18.05cm]{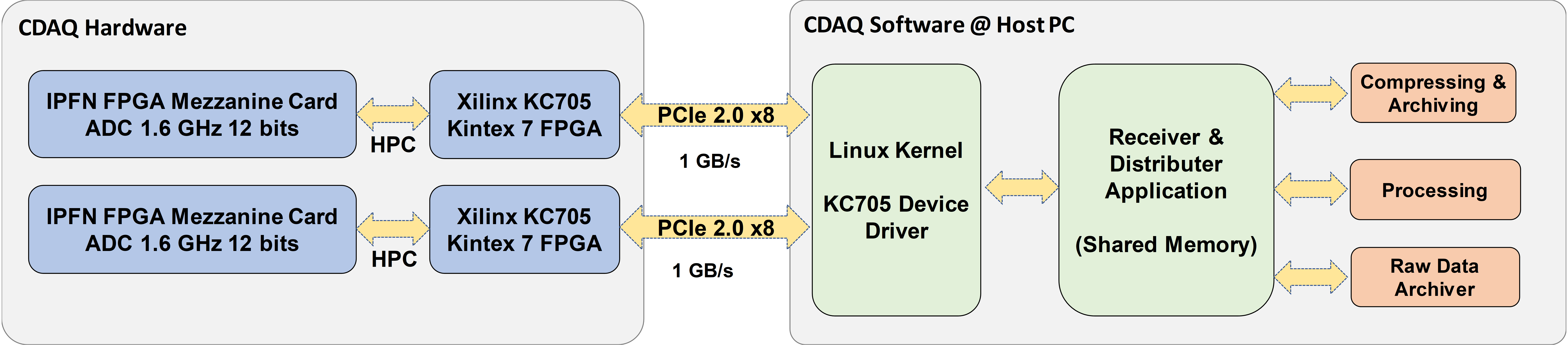}
	\caption{Global Real-time Software Architecture}
	\label{fig:Architecture}
\end{figure*}

This section depicts the design of the real time architecture to acquire the pulse information from the detectors and build the particles spectra using pulse integration, neutron/gamma separation and pile-up correction algorithms. The inversion algorithm to calculate the emissivity profile will run in a different dedicated host connected through a dedicated Ethernet link. Figure \ref{fig:Architecture} shows the software architecture of the present RNC prototype.

Figure \ref{fig:Architecture} also depicts the data links and hierarchy between the main software modules:
\begin{itemize}
	\item Linux device driver;
	\item Data receiver and distribution;
	\item Pulse processing for particle energy spectrum construction;
	\item Data compressing and archiving;
	\item Raw data archiver (for testing and validation).
\end{itemize}

The use of a shared memory permits the distribution of the pulse data among several clients in real time so that the most recent pulse data is available for different purposes. The pulse processing application calculates the particle energy spectrum for real time measurement, while the data compression \cite{IEEERT2018:Bruno} is used to store or deliver to the archiving system the raw pulse data for offline analysis and physics studies. The raw pulse data archiver may be used when no data compression is desired or for testing and validation activities. 

To optimize the performance and the real time deterministic completion of the tasks, each software module runs in dedicated isolated cores. The load of each CPU core was measured during preliminary tests to help dimensioning the distribution of tasks per CPU core and the number of CPU cores per task. For the present prototype tests the following needs were identified for optimal performance at 2 Mevents/s:
\begin{itemize}
	\item 2 logical cores for the use of the Operating System;
	\item 1 logical core per KC705 for the use of the Device Driver;
	\item 1 logical core per KC705 for the use of the Data Receiver and Distribution application;
	\item 3 logical cores per channel for the use of the Pulse Processing Algorithm; 
	\item 5 logical cores per KC705 for Data Compression Algorithm \cite{IEEERT2018:Bruno}.
\end{itemize}

\begin{figure}
	\centering
	\includegraphics[width=8.05cm]{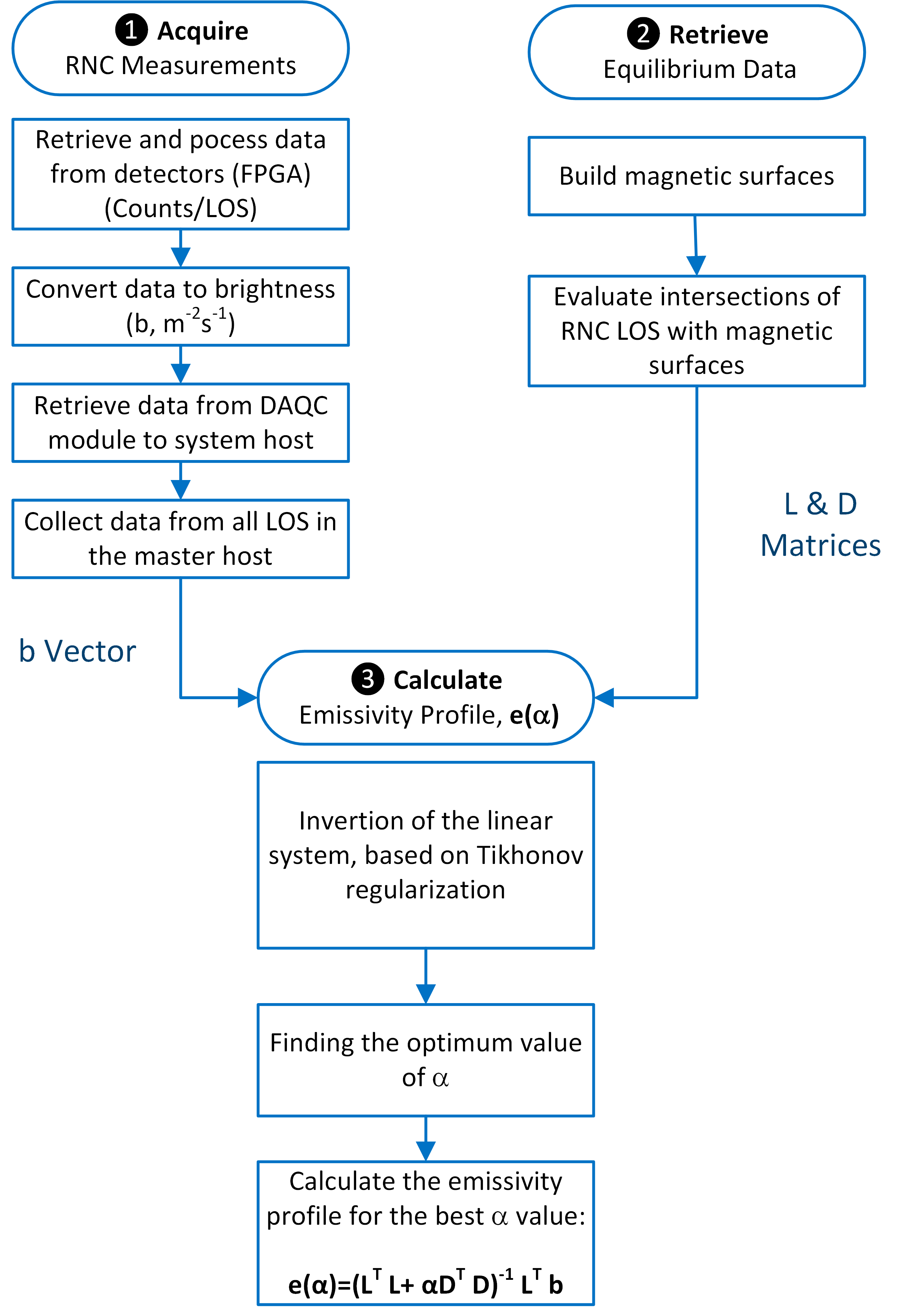}
	\caption{Processing tasks for the neutron emissivity reconstruction algorithm based on Tikhonov regularization method.}
	\label{fig:Flowchart_Emissivity_Reconstruction}
\end{figure}

Figure \ref{fig:Flowchart_Emissivity_Reconstruction} depicts the processing tasks breakdown structure identified to obtain the real time measurement of the neutron emissivity profile \cite{RNC_Marocco_2016}\cite{RNC_Riva_2018}:
\begin{itemize}
	\item Task 1 - Acquire and process neutron detector pulses, building the necessary energy spectrum for neutrons and gammas;
	\item Task 2 - Retrieve and calculate the necessary inputs from the plasma equilibrium data;
	\item Task 3 - Calculate the neutron emissivity profile using the inputs from the previous tasks and running an inversion algorithm based on Tikhonov regularization method \cite{Marocco_2011}. 
\end{itemize}
 
The identified tasks were then distributed in the real time software architecture design that implements the parallelization of the algorithms by 
(i) running the neutron count detection continuously within the control cycle time slots; (ii) calculating the emissivity profile using the acquired data in the previous cycle and sending the data to the real time control network within the same control cycle slot; (iii) performing the real time measurement control cycle in 10 ms.

\begin{figure*}
	\centering
	\includegraphics[width=16cm]{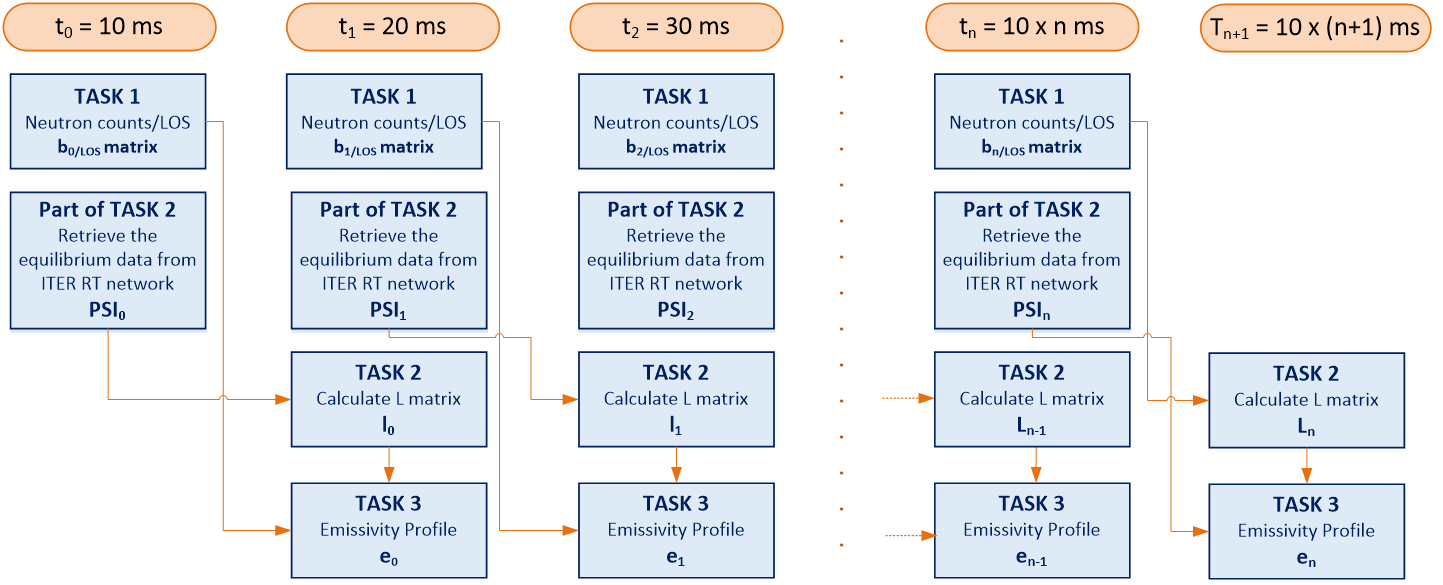}
	\caption{Neutron emissivity profile control cycle diagram.}
	\label{fig:ControlCycle}
\end{figure*}

Figure  \ref{fig:ControlCycle} outlines the real time measurement control cycle diagram with the appropriate distribution of tasks to perform the control cycle under 10 ms:
\begin{itemize}
	\item Task 1 runs immediately after acquisition starts, before any other task, for the complete period of a control cycle;
	\item Task 2a retrieves the plasma equilibrium data necessary to reconstruct the magnetic flux surfaces; 
	\item Task 2b uses data retrieved in the previous control cycle  and calculates the necessary inputs from the plasma equilibrium data;
	\item Task 3 calculates the neutron emissivity profile using the neutron and magnetic flux data available from the previous cycle as well as calculations made in the present control cycle;
	\item The tasks run in parallel using different CPUs for improved performance and independent run time processing.
\end{itemize}

\subsection{Data Distribution}

The host computer receives the pulse data using Direct Memory Access (DMA) channels programmed by the host and managed in the FPGA. The DMA transfer engine may generate hardware interrupts to signal the host the data is available. The developed Linux PCIe driver makes the received data from the hardware available to the data distribution application by storing it in an internal memory. On top of the driver a data distribution application prepares the data in a shared memory using a round buffer. The data can be accessed by client applications using the data pointers managed by the distribution applications, that logs any data lost between the client application and the distribution application. 

Although the usage of the polling mechanisms in the driver are not the usual approach for variable event incoming data rate, the tests using the interrupt mechanism reflected a poor stability due to errors in the interrupt packets transmission when the maximum event rate is used. An architecture using a different device driver approach with polling mechanism and an internal data transmission correction algorithm  was implemented, aiming at improving the performance and reliability. This enables the device driver to automatically check and recover missing data blocks, recovering the data losses in a transparent way for the client applications \cite{Soft2018_Bsantos}. 

\subsection{Data Compression }

Besides the real time measurements for control purposes, the RNC diagnostic data acquisition system is also required to send to the ITER data archiving the pulse data with a sustained 2 Mevent/s peak event rate, producing up to
0.5 GB/s of data per channel. 

To minimize the data size to transfer to the data archiving system a lossless compression algorithm was implemented, providing
high compression speeds to comply with the system requirements. During the RNC diagnostic prototype phase, the compression algorithm was tested using data across multiple processor cores with good results \cite{IEEERT2018:Bruno}, which are important for dimensioning the system in the final design phase.  

\subsection{Pulse Processing}

\begin{figure}
	\centering
	\includegraphics[width=5.85cm]{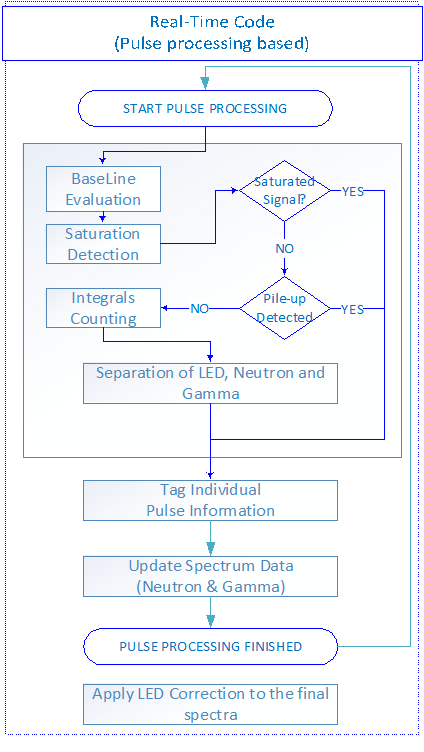}
	\caption{Pulse processing real-time code software structure.}
	\label{fig:Flowchart_Pulse_Processing}
\end{figure}

The pulse processing algorithm aims at delivering the neutron and gamma energy spectra by processing the pulse data acquired from the detectors for each LOS. Figure \ref{fig:Flowchart_Pulse_Processing} depicts the real time pulse processing algorithm flowchart, including the main functions:
\begin{itemize}
	\item Baseline evaluation;
	\item Saturation detection;
	\item Pile-up detection;
	\item Signal integration (energy calculation);
	\item Signal and particle separation;
	\item Construction of energy spectrum per particle.
\end{itemize}

For real time energy calibration it is foreseen the use of a known energy LED. The application of LED correction at the end of the control cycle to calibrate the energy spectra in real time is under evaluation. In former offline pulse processing implementations the LED correction was applied to each pulse individually, however to improve the performance of the pulse processing algorithm the LED correction may be applied only to the final spectra. This has implications in the resolution and may introduce a deviation of the real spectrum that must be validated. Another solution envisaged is to apply the correction to every pulse based on the LED energy detected on the previous control cycle. This will minimize the implications presented with a small increase in the processing time. Moreover, the stability of the offset variation is enough to permit that the correction is made using the LED correction of the previous 10 ms cycle.

The number of pile-up and saturated signals detected are used as a correction factor for statistical spectrum count correction. The investigation of pile-up separation algorithms in FPGA and host computer is foreseen for the future, but present results with the statistical correction were validated.

\subsection{Neutron Emissivity Profile Algorithm}

The ITER neutron emissivity profile is calculated using an inversion algorithm  applying the Tikhonov regularization method \cite{Marocco_2011}. 

Figure \ref{fig:RNC_LOS} shows the RNC geometry composed by (i) lower and higher plasma inspection LOS using in-port detectors; (ii) centre plasma inspection LOS using ex-port detectors. The neutron emissivity profile is calculated in the intersection points of the different LOS and the magnetic flux surfaces. 

Figure \ref{fig:Flowchart_Emissivity_Reconstruction} describes the algorithm used for the complete calculation. The RNC measurements ($b\_signal$) and the L matrix ($L-Matrix$) are obtained in tasks 1 and 2, while the regularization matrix ($D$) is pre-calculated based on the system geometry. In task 3 the inversion algorithm to obtain 2D neutron emissivity profile is performed \cite{Marocco_2011}.

\section{Validation and Performance Tests}

This section details the results of the algorithm validation and performance tests for the more critical algorithms and global prototype architecture. 

\subsection{Pulse Processing}

\begin{figure}
	\centering
	\includegraphics[width=6.50cm]{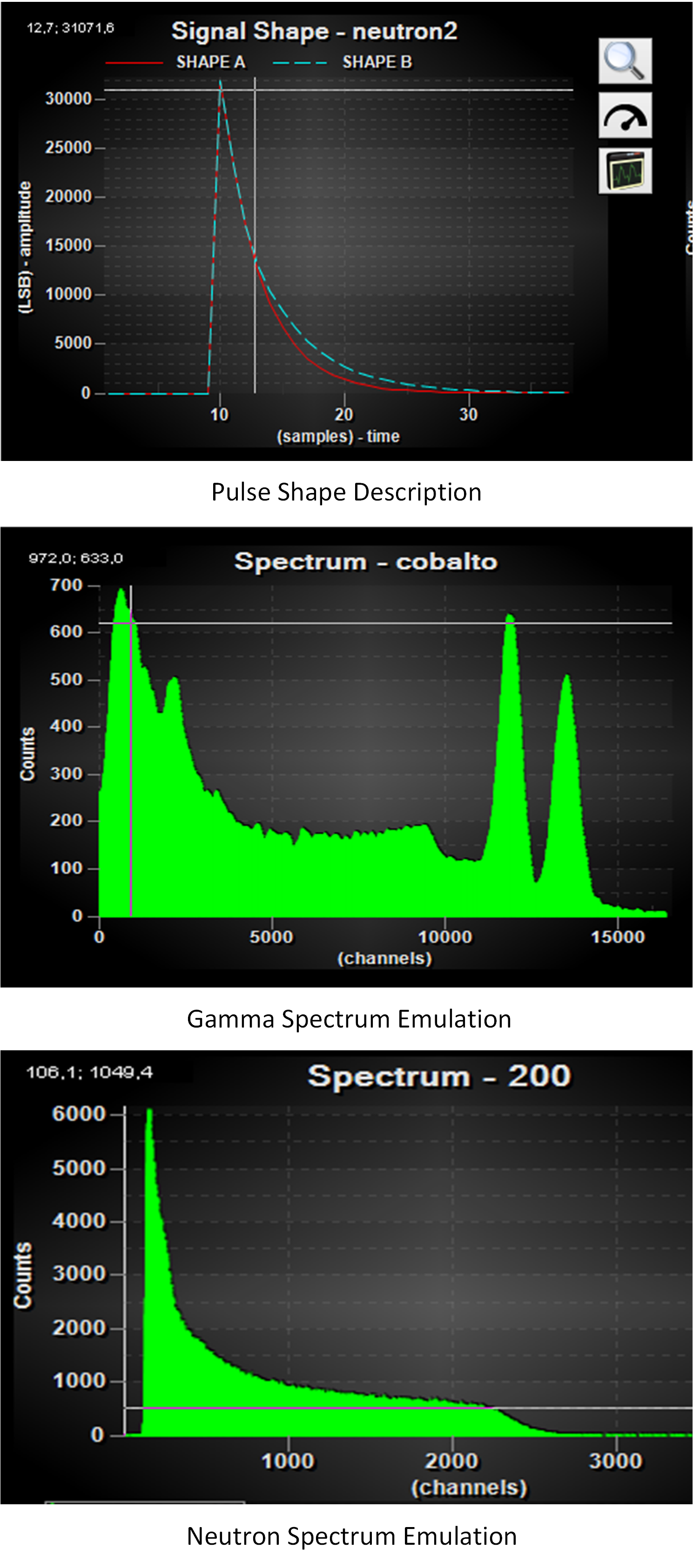}
	\caption{Pulse emulator information to generate the simulated neutron and gamma spectra.}
	\label{fig:Pulse_Generator}
\end{figure}

To validate the pulse processing algorithm the 2 channel generator CAEN DT5800D  was used. Each channel was programmed with exponential decay signal with different decay times aiming at emulating neutron and gamma particle detector signal. A different spectrum was programmed in each channel output. Figure \ref{fig:Pulse_Generator} depicts the signal shapes and the corresponding spectrum for neutron and gamma particles. Both signal generator output channels were mixed with Mini-Circuits ZX10R-14-S+, allowing signal bandwidth from DC to 10 GHz. This signal is then fed into the system FMC analogue input to be acquired and processed by both algorithms in the FPGA firmware and in the host PC.

\begin{figure}
	\centering
	\includegraphics[width=7.85cm]{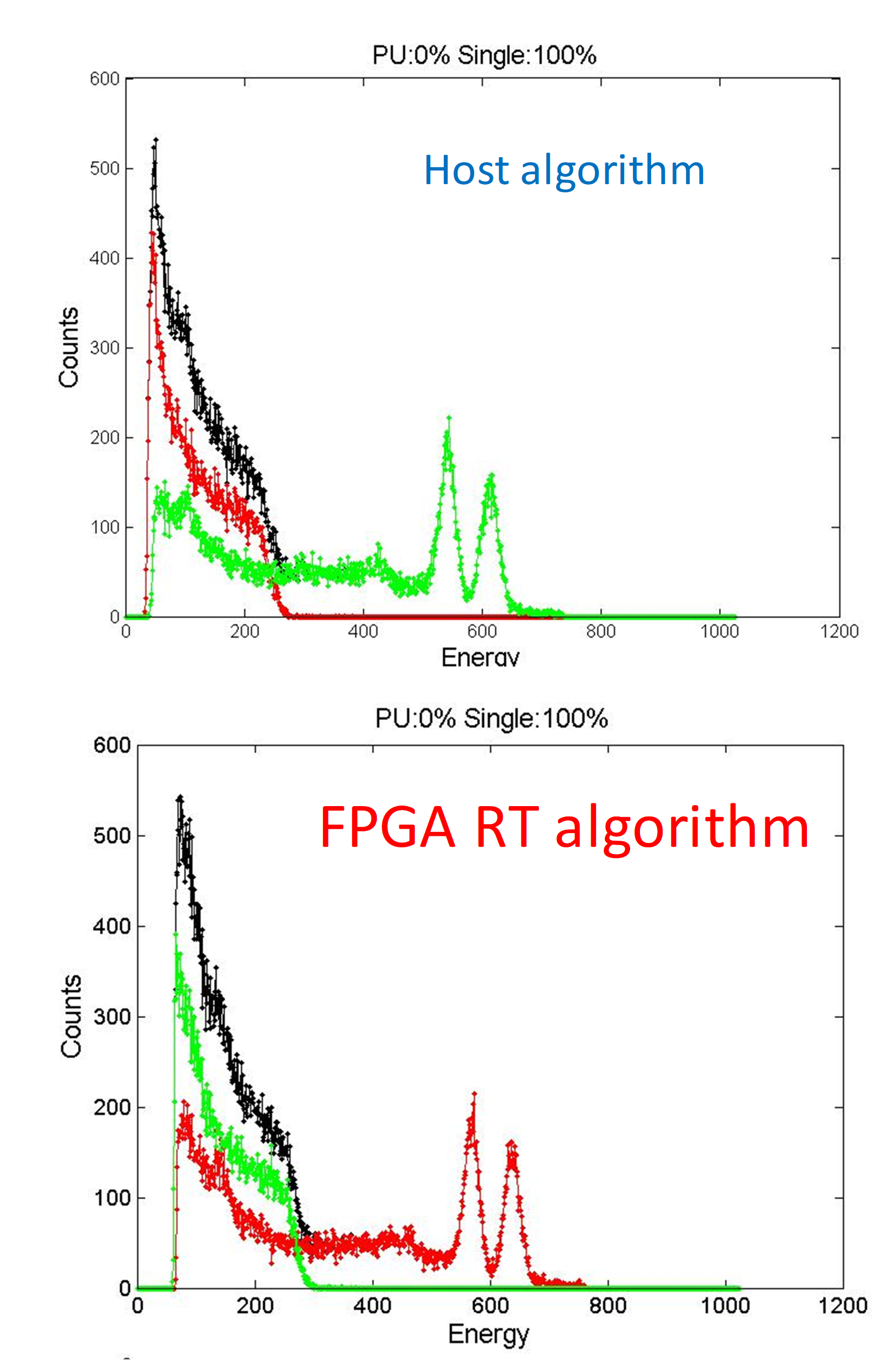}
	\caption{Cross validation between host and FPGA algorithms for neutron and gamma spectra pulse discrimination.}
	\label{fig:Particle_Descrimination}
\end{figure}

The algorithm validation was made by comparing the known input spectra signals with the output spectra from the FPGA algorithm and host algorithm. Figure \ref{fig:Particle_Descrimination} shows the spectra outputs for both algorithms that are in agreement with each other and moreover with the known inputs from the signal generator.  

\begin{figure}
	\centering
	\includegraphics[width=7.85cm]{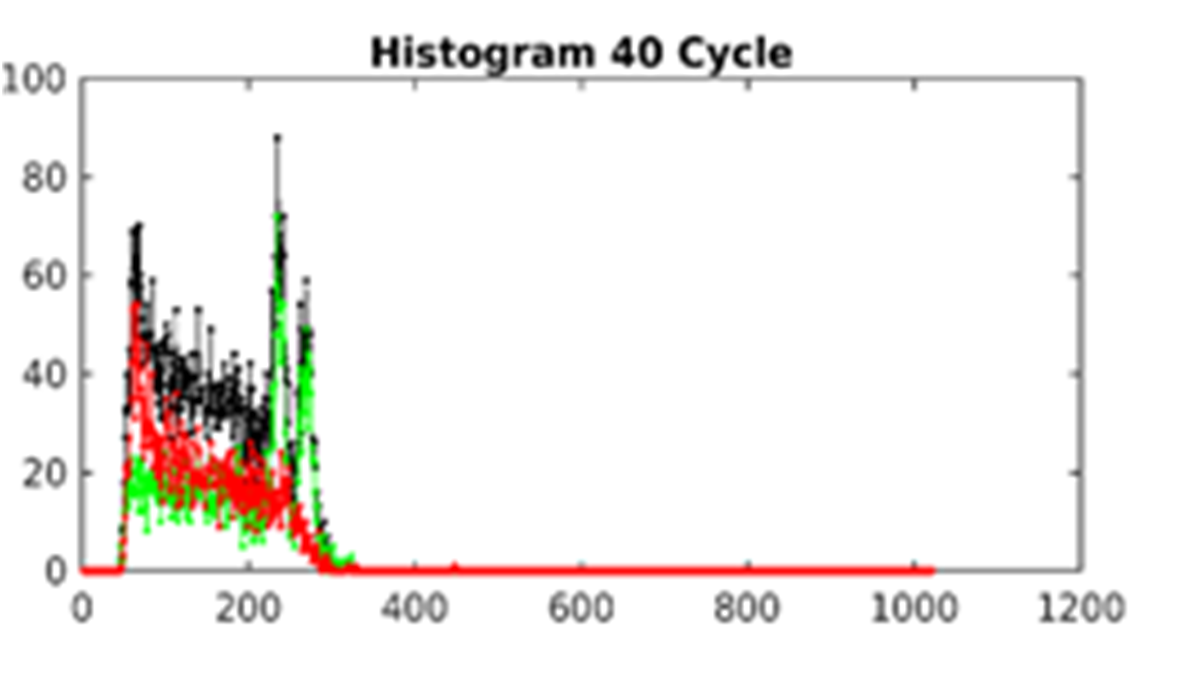}
	\caption{Spectrum obtained for 1.5 Mevents/s with 10 ms of data with no data loss.}
	\label{fig:Pulse_Processing_10ms}
\end{figure}

The signal generator was also programmed to feed the prototype system with different  pulse event rates aiming at measuring the performance of the host algorithm. The processing time for 3 different event rates (2.0 Mevents/s, 
1.5 Mevents/s and 1.0 Mevents/s) using 1 logical core with 10 ms control cycle running the complete data analysis was measured during several cycles. Figure \ref{fig:Pulse_Processing_10ms} presents the typical spectrum algorithm output for 1.5 Mevents/s. 

\begin{figure}
	\centering
	\includegraphics[width=9.05cm]{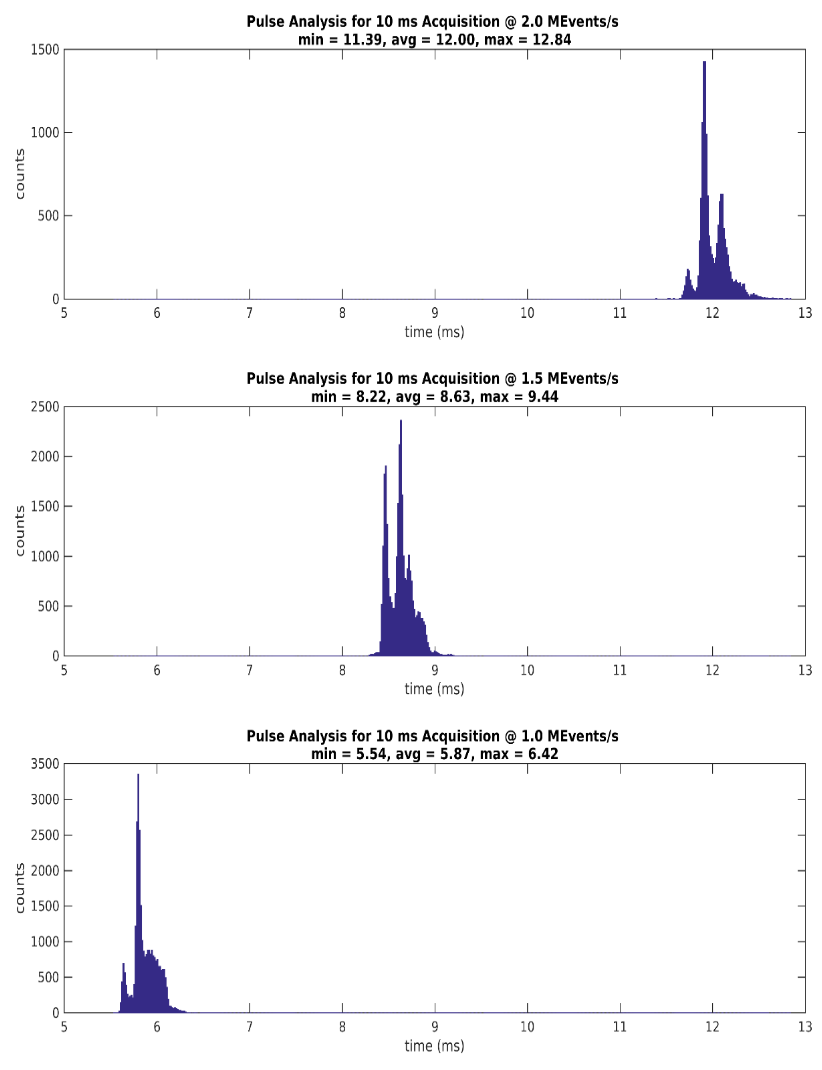}
	\caption{Performance of the pulse processing algorithm. Run time calculation for (i) 2.0 MEvents/s; (ii) 1.5 MEvents/s and (iii) 1.0 MEvents/s.}
	\label{fig:Performance_Pulse_Processing}
\end{figure}

Figure \ref{fig:Performance_Pulse_Processing} depicts the performance results by plotting the processing times during several control cycles for 2.0, 1.5 and 1.0 Mevents/s. It is possible to see that for 1.0 and 1.5 Mevents/s processing times were always under 10 ms, which is in the limit to comply with the specification for the system processing time. For 2.0 Mevent/s however all cycles needed more than 10 ms to process the data, which is out of the system specification. It is necessary to upgrade the implementation to achieve 2.0 Mevents/s processing time under 10 ms using the host algorithm by allocating 2 logical cores for parallel pulse processing.

\subsection{Neutron Emissivity Profile Algorithm}
\begin{figure*}
	\centering
	\includegraphics[width=17.85cm]{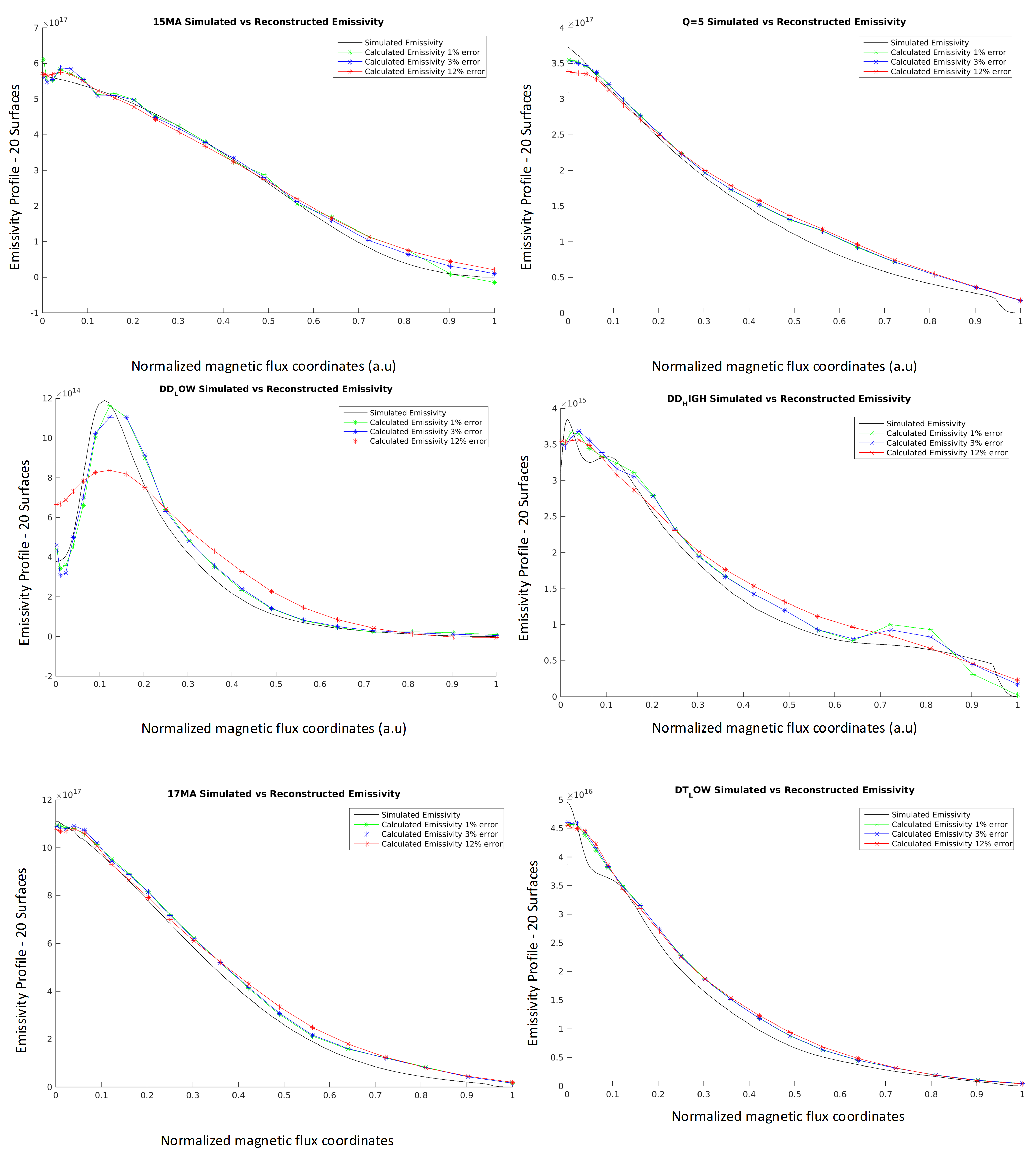}
	\caption{Validation of ITER plasma scenarios reconstructed emissivity for different levels of random noise using 20 magnetic surfaces (from left to right, top to bottom): (i) 15MA scenario; (ii) Q=5 scenario; (iii) DD-Low scenario; (iv) DD-High; (v) 17MA scenario; (vi) DT-Low scenario.}
	\label{fig:ITER_Scenarios_Emissivity}
\end{figure*}

The validation of the neutron emissivity reconstruction algorithm was performed to understand if the code runs and converge with the correspondent simulated emissivity for relevant ITER scenarios. A cross-check of  the reconstructed emissivity, for different level of random noise (1\%, 3\% and 12\%) using 20 magnetic surfaces, with the correspondent simulated emissivity was done for the relevant ITER scenarios. Figure \ref{fig:ITER_Scenarios_Emissivity} depicts the results of this validation. The only noticeable difference that was verified is for 12\% input error in the DD-LOW scenario. 

\begin{figure}
	\centering
	\includegraphics[width=9.05cm]{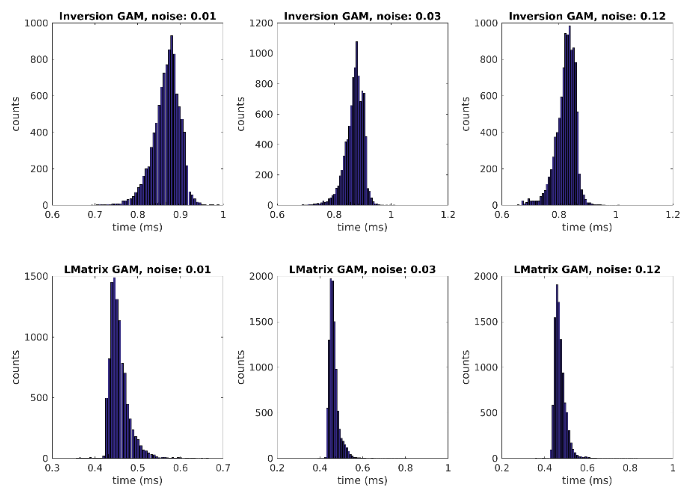}
	\caption{Run time calculation results for the two processing modules of the real-time inversion code using inputs with different level of random noise (1\%, 3\%, 12\%) and considering 20 magnetic surfaces. Top: results for the inversion GAM. Bottom: Results for the LMatrix GAM.}
	\label{fig:Performance_Inversion_Algorithm}
\end{figure}

Figure \ref{fig:Performance_Inversion_Algorithm} shows the processing times for the neutron emissivity profile for measurement data with different levels of random noise. It is possible to confirm that the complete task runs always under 1.5 ms which is compliant with the system specification.

\section{Summary and Future Work}

This section presents the prototyping activity most relevant achievements and important conclusions that can be taken for the improvement of the final system design. 

A system architecture has been presented to help the final RNC system design and specifications.The system prototype has been used to:
			\begin{itemize}
				\item Implement the most critical algorithms;
				\item Validate the results of the algorithms;
				\item Measure the performance of the system.
			\end{itemize}
			
The performance tests were relevant to help retrieve valuable information to size the optimal system configuration in terms of processing needs and number of acquisition channels per CPU. 

The inversion algorithm to calculate the  neutron emissivity profile can run in less than 2 ms which is fully compliant with the system specifications.

On the contrary, the pulse processing algorithm runs in 12 ms using 1 CPU core for peak event rate of 2.0 Mevents/s, which is not compliant with the system specifications. Nevertheless, the use of more CPU cores to parallelize the host algorithm is one solution that may be implemented in the current prototype to fulfill the system requirements. However, the FPGA pulse processing algorithm already implemented and running in compliance with the timing system specification should be used to improve the performance of the system \cite{IEEERT2018:Ana}.

Currently tests with the prototype system using two FPGA Xilinx modules in the same host, providing the host with 4 acquisition channels to acquire and process in real time are ongoing.


%

\section*{Acknowledgment}

The work leading to this publication has been funded partially by Fusion for Energy under the Contract F4E-FPA-327. IST activities also received financial support from "Funda\c{c}\~ao para a Ci\^encia e Tecnologia" through project UID/FIS/50010/2013. This publication reflects the views only of the author, and Fusion for Energy cannot be held responsible for any use which may be made of the information contained therein.

This manuscript is in memory of Professor Carlos Correia
who is no longer among us.

%







\end{document}